\documentstyle[sprocl,epsfig]{article}
\newcommand{\be}{ \begin{eqnarray}}
\newcommand{\ee}{\end{eqnarray}}
\newcommand{\beno}{ \begin{eqnarray*}}
\newcommand{ \eeno}{\end{eqnarray*}}
\newcommand{\bfg}{\begin{figure}}
\newcommand{\efg}{\end{figure}}

\begin{document}

\title{The problem of Chiral Restoration \\  
 and Dilepton Production in Heavy Ion Collisions }
\author{E.V.Shuryak}

\address{Department of
  Physics and Astronomy, \\
  State University of New
  York, \\ Stony Brook NY 11794  USA\\E-mail: shuryak@dau.physics.sunysb.edu} 

\maketitle
\abstracts{
%\centerline{\it From the Plank Length to Hubble Radius, Erice, Sept. 98 }
%\vskip 1cm
In the lecture we review several issues
related to recent development in non-perturbative
  QCD.   The
  ``instanton liquid  model''  reproduces not only the basic vacuum
  parameters (the condensates) but even hadronic correlators. New
   information obtained from lattice simulations also confirm
  it. Meanwhile the model itself was developed into a self-consistent approach,
  allowing to include  't Hooft interaction  to all orders.  It  was
also  generalized to non-zero temperatures and high densities. 
We discuss one issue, displayed by behavior of the pion and rho
correlation functions: the former has strong non-perturbative effects
at small distances, the latter has none. What happens
at $T\sim T_c$ can be answered by dilepton production experiments with
 heavy ion collisions. The results definitely indicate large changes
in spectral density and ``melting" of the rho, possibly with
reaching chiral restoration.
}

\section{Chiral symmetry and instantons in vacuum/hadronic structure}
 Let me start the lecture reminding few well-known facts from the textbooks.
 If quark masses are ignored, the fermion part of
the QCD Lagrangian becomes a sum of
two  independent terms, with left and right-handed quarks. The possibility
to rotate those in flavor space $independently$ generates {\it two} additional
  ``chiral" symmetries, $U(1)_A$ and $SU(N_f)$ ones,
which  have rather different fate.

 The $U(1)_A$ one  (generated by $exp(i\phi\gamma_5)$ rotation)
is explicitly broken by the so called chiral anomaly, and so at quantum level
it is
simply $not$ a symmetry of QCD.
The strength of its violation can be seen from a 
deviation of the pseudoscalar singlet $\eta'$ mass (959 MeV)
 from that of a pion/kaon/eta multiplet: note that it is  surprisingly
large.

 The $SU(N_f)$ part of the chiral symmetry
 is {\it spontaneously broken},  the QCD vacuum  is asymmetric.
Its measure is the so called quark
  condensate
$<\bar q q>$. By Goldstone theorem, 
 massless modes (rotations to other equivalent vacua) appear, which are pions.
General features of their interactions are described by chiral effective Lagrangians.

  The main questions we are going to discuss are
related to the {\it underlying  dynamics} of these
 phenomena (and those are rarely discussed in textbooks).
 However the earliest  
attempt to explain  chiral symmetry breaking 
was made as early as 1961 \cite{NJL}.  By analogy to
superconductivity, it was shown that sufficiently strong  
attraction
between quark and antiquark in the scalar channel can re-arrange the vacuum,
create the quark condensate and a ``gap" at the surface of the Dirac sea,
the quark effective mass. I would argue below that it is
 a correct idea, and that understanding
of the origin of that attractive interaction and its
exact form in QCD was clarified only during the last
two decade  (see e.g. references in a review \cite{SS_98}).

   What was found (first empirically, then from the
success of the so called instanton
liquid models, then from lattice studies) is that both explicit
breaking of U(1) and spontaneous
breaking of $SU(N_f)$ $SU(N_f)$ chiral symmetries are driven by 
 instantons. 

The first part was easier to understand:
as G.t'Hooft have explained in 1976, instantons generate $2*N_f$-fermion
effective vertices of particular structure, violating U(1) by flipping
quark chiralities. Quantitative part of the U(1) problem was later
related to the so called ``topological susceptibility" by 
Witten and Veneziano, and recent lattice studies 
have left no doubts that those are indeed
completely saturated by well-identified instantons. 

    However spontaneous breaking is not seen at the level of one instanton,
one need some knowledge about their
ensemble, and particular conditions should be met for it to occur.
Otherwise (and this happens at high enough temperature T or large enough 
number of quark flavors $N_f$) the chiral symmetry remains unbroken.
In 1982 I have fixed the mean density and size of the instantons
\cite{Shu_82} to be
\be
\label{param}
n=n_+ + n_- \approx 1 fm^{-4}; \,\,\, \rho\approx 1/3 fm
\ee
It is still rather dilute ensemble because $n\rho^4\sim 1/81$, but it
is dense enough to be in the chirally broke phase! 

\begin{figure}[h]
\begin{center}
\leavevmode
\epsfig{file=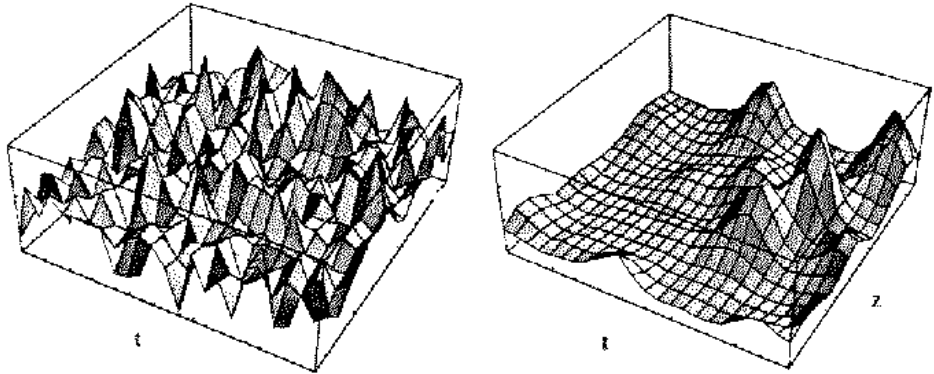, width=5in}
\end{center}\caption{\label{cooling}
Sample of a  gauge configuration before (left) and after
  (right) cooling.In the latter case instantons are clearly seen.
 The quantity shown is the action density, and its scale (not shown) is
 two orders of magnitude larger on the left figure.
}\end{figure}

A decade later lattice configurations were stripped of the 
 ``fog'' of quantum fluctuations and their  classical content was revealed.
 In Fig.\ref{cooling} from \cite{CGHN} one can see how it works, so that 
one can see and count instantons. The values given  in (\ref{param})
were basically confirmed.
%\widetext
% figure I used

 Instantons do all what the hypothetical NJL interaction was 
supposed to do. Moreover, 
they do this job better, because they generate vertices with the 
particular non-locality:
therefore  nasty questions 
related with the non-renormalizable NJL model are naturally  resolved, and the
NJL cut-off $\Lambda_{NJL} \sim 1 GeV$ (also known as ``the chiral scale")
was attributed to the instanton sizes. 
Furthermore, a  practical approach was discovered (the ``interacting instanton
liquid model", or IILM) 
 allowing to 
include {\em all} orders in the instanton-induced ('t Hooft) effective 
interaction.

  But in the last few years it became more and more clear
that instantons are responsible for nearly all\footnote{ 
With the notorious exception of confinement.}
non-perturbative phenomena associated with light quarks, their
propagation in vacuum and bound states. Their masses are mostly the 
masses of ``constituent quarks" we already mentioned, and even their
spin-dependent forces seem to be instanton-generated as well\cite{spin}. There are
direct indications, that not only pions but even the usual mesons like
$\rho$ and baryons like nucleon are in a way collective excitations of
the chiral condensate.

Finally, to complete the overview of instantons,
 let me mention that recent progress in
supersymmetric gauge theories have indicated some surprising things about
them as well.
In particular, partial exact solution for $N=2$ 
SQCD due to Seiberg and Witten\cite{SW_94})
can be  expanded in inverse powers of small parameter $\Lambda/a$ (where $a$ is the Higgs VEV)  at large $a$, one can 
see that (apart of a single perturbative one-loop log) all the power terms are
 ($\Lambda/a)^{4*integer}$,
just like multi-instanton contributions should give.
The first two orders have been 
calculated\cite{susy_inst} and were found to be exactly right. Although 
higher orders 
are not yet done, it is highly possible that that in this theory 
instantons are the {\em only} dynamics there is, in the sense that summing the
series in their interactions (an analog of the instanton liquid calculation
we will discuss below) provides the $exact$ result.

  The amusing similarity between QCD and (its relative) the N=2 SQCD have been
recently demonstrated in \cite{RRS}. It is related to the issue of
already mentioned ``chiral scale" 1 GeV. In QCD it is phenomenologically
known that this scale is not only the upper bound of effective theory but also
the lower bound on parton model description. However, one cannot really see it
from the perturbative logs: 1 GeV is several times larger than their 
natural scale, $\Lambda_{QCD}\sim 200 MeV$. In the N=2 SQCD the answer is known:
effective theory at small $a$ (known also as ``magnetic" formulation) is separated
from perturbative region of large $a$ by a singularity, at which monopoles
become massless and also the effective charge blows up. How it happens also  
follows from Seiberg-Witten solution, see Fig.2. Basically the perturbative
log becomes cancelled by instanton effects,  long before the charge blows
up due to ``Landau pole" at $p\sim \Lambda$. It happens ``suddenly" because
instanton terms have strong dependence on $a$: therefore perturbative
analysis seems good nearly till this point.

For comparison, in QCD we have calculated effective charge with the instanton correction, as
defined by Callan-Dashen-Gross expression. All we did was to
put into it the present-day knowledge of the
instanton density. The resulting curve is astonishingly similar to
the one-instanton one in N=2 SQCD. Note, that  in this case as well,
 the ``suddenly
appearing" instanton effect blows up the charge, making perturbation
theory inapplicable, and producing massless pions, the QCD ``magnetic"
objects.  Moreover, it even happens at about the same place! (Which is
probably a coincidence.) 

\begin{figure}[t]
\vskip -0.4in
\epsfxsize=3.8in
\centerline{\epsffile{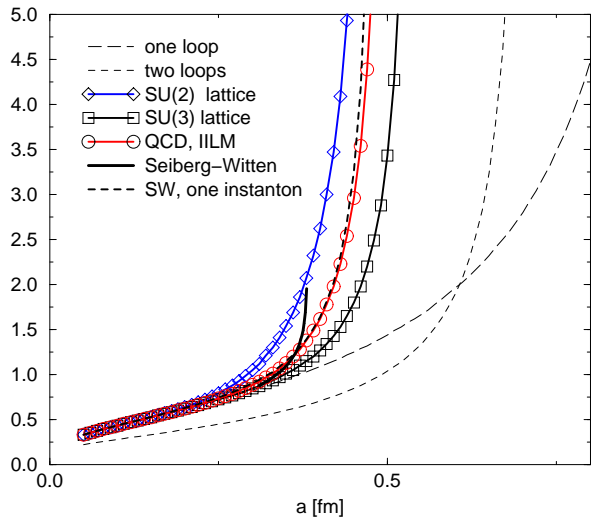}}
\vskip -0.05in
\caption[]{
 \label{figure1}
 The effective charge $b \,g^2_{eff}(\mu)/8\pi^2$ (b is the coefficient
of the one-loop beta function) versus normalization scale $\mu$ (in units of
its value at which the one-loop charge blows up). The thick solid line
correspond to exact solution \cite{SW_94} for the N=2 SYM, the thick dashed line
shows the one-instanton correction. Lines with symbols (as indicated on figure)
stand for N=0 QCD-like theories,
SU(2) and SU(3) pure gauge ones and QCD itself. Thin long-dashed and short-dashed lines are one and two-loop results.
}
\end{figure}

The behavior is shown in  Fig.\ref{figure1}, where we
have included both a curve which shows the full 
coupling (thick solid line), as well as a curve which illustrates
only the one-instanton correction (thick dashed one). Because we will want
to compare the running of the coupling in different theories,
we have plotted $b g^2/8 \pi^2$ (b=4 in this case is the one-loop coefficient
of the beta function) and measure all quantities 
in  units of $\Lambda$, so that  the one-loop
charge blows out at 1. The meaning of the scale
can therefore be determined by what enters in the logarithm.

  Recent conjectured correspondence between N=4 SQCD and string theory/SUGRA
in 5d anti-de Sitter ($AdS_5$) space (discussed here in  lecture by E.Witten) 
has also an astonishing instanton connection \cite{Mattis_etal}.
 In the large $N_c$ limit, and in
the weak coupling domain, the results are dominated by specific 
``multi-instanton molecules" in which all instantons are at the the same 
point and have the same sizes\footnote{ They fit there without a problem due to 
 the limit of very large 
number of colors.}. This is an example of a (long-predicted) ``master field".
Any Green function look like propagators going from various space-time points to
the point in $AdS_5$, which happen to be nothing else but the instanton 5 coordinates $d^4z d\rho/\rho^5$. Remarkably, even another 5-d sphere appears, from ``di-fermion" condensates, and so the result looks in truly amazing correspondence with the
``holographic principle" Witten spoke about.

In summary,
   instantons are responsible for a variety of
non-perturbative phenomena, and all of us should
study them more. 

\section{Phenomenology of QCD correlation functions}

  In this section we go from very exciting recent results advertised above
back to simple observations from which the first arguments about a prominent
role of instanton-induced effects in QCD were first deduced.

In confining gauge theory like QCD the correlation functions of (gauge invariant) 
local operators are the best possible tool to bridge the gap between
the fundamental fields and physical excitations. 
 The same  functions can be calculated at large distances x
using  the physical states 
(mesons, baryons, glueballs), and at small x  in terms of quarks 
and gluons. Some of them can be completely deduced from experimental data
(see below) and all of them can be obtained from lattice simulations.

Loosely speaking, hadronic correlation functions play the same role for 
understanding the forces between quarks as the $NN$ scattering phase did in 
the case of nuclear forces. In the case of quarks, however, confinement 
implies that we cannot define scattering amplitudes in the usual way. Instead, 
one has to focus on the behavior of gauge invariant correlation functions at 
short and intermediate distance scales. The available theoretical and 
phenomenological information about these functions was  reviewed 
in\cite{Shu_93}. 
 
Euclidean point-to-point correlation functions are defined as 
\be
\label{cor_gen}
 \Pi_h(x) &=& \langle 0|j_h(x)j_h(0)|0\rangle  ,
\ee
where $j_h(x)$ is a local operator with the quantum numbers of a hadronic 
state $h$.
Hadronic correlation functions can be written in terms of the spectrum and the 
coupling constants of the physical excitations with the quantum numbers of the 
current $j_h$. This connection is based on the standard dispersion relation
\be 
\label{disp_rel}
 \Pi(Q^2) &=& \frac{1}{\pi} \int ds\,
  \frac{{\rm Im}\Pi(s)}{s+Q^2} + a_0 + a_1 Q^2 + \ldots ,
\ee
(where $Q^2=-q^2$ is the Euclidean momentum transfer and we have indicated 
possible subtraction constants $a_i$) and the spectral decomposition 
($\rho(s)\equiv\frac{1}{\pi}{\rm Im}\Pi(s)$)
\be
\label{spec_rep}  
\rho(s=-q^2) &=& (2\pi)^3 \sum_n \delta^4(q-q_n) 
  \langle 0|j_h(0)|n\rangle \langle n|j_h^\dagger(0)|0\rangle . 
\ee
A spectral representation of the coordinate space correlation function
is obtained by Fourier transforming (\ref{disp_rel}), 
\be
\label{coord_rep}
 \Pi(\tau) &=& \int ds\,\rho(s) D(\sqrt{s},\tau),
\ee
where $D(m,\tau)=m K_1(m\tau)/(4\pi^2\tau)$ is the Euclidean propagator of a 
scalar particle with mass $m$. Note that for large arguments the correlation 
function decays exponentially, $\Pi(\tau)\sim \exp (-m\tau)$, where the decay 
is governed by the lowest pole in the spectral function. 

Correlation functions that involve quarks fields only 
 can be expressed in terms of the  quark 
propagator. For an isovector meson current $j_{I=1}=\bar u\Gamma d$ (where 
$\Gamma$ is only a Dirac matrix), the correlator only has a ``one-loop'' 
contribution ($S^{ab}(x,y)$ is the quark propagator)
\be
\label{mes_cor}
 \Pi_{I=1}(x) &=& \langle {\rm Tr}\left[ S^{ab}(0,x)\Gamma
 S^{ba}(x,0)\Gamma \right] \rangle .
\ee
The averaging should be performed over all gauge configurations, with proper
weight function $\det(iD\!\!\!\!/\,+im)\exp(-S)$. Correlators of isosinglet
meson currents $j_{I=0}=\frac{1}{\sqrt{2}}(\bar u\Gamma u+\bar d\Gamma d)$ 
receive an additional two-loop, or disconnected, contribution
\be
\label{sing_cor}
 \Pi_{I=0}(\tau) &=& \langle {\rm Tr}\left[ S^{ab}(0,x)\Gamma
 S^{ba}(x,0)\Gamma \right] \rangle  - \\
  && \hspace{1cm} 2 \langle {\rm Tr}\left[ S^{aa}(0,0)\Gamma
 \right]\, {\rm Tr}\left[ S^{bb}(x,x)\Gamma \right]\rangle .
\nonumber
\ee
Analogously, baryon correlators can be expressed as vacuum averages of three 
quark propagators. 

 In this lectures there is no time to discuss this subject in details
( see review \cite{Shu_93}), and so we only discuss the amazing difference
between the vector ($\rho$) and the 
pseudoscalar channels ($\pi,\eta'$) observed
experimentally. Then we would show how instantons explain this behavior, at least
qualitatively. Of course, vector channel is also of special importance because later
on we would move to discuss its modifications at high temperatures, as revealed
by dilepton production in heavy ion collisions.

In the case of the vector-isovector 
channel the data from $\sigma (e^+e^-\to (I=1~{\rm hadrons}))$ and
$\tau$-lepton decay\footnote{Those give also the axial-vector $a_1$ 
spectral density,
 from hadronic decays of the $\tau$ lepton, $\Gamma(\tau\to\nu_\tau 
+{\rm hadrons})$.} . 

\begin{figure}
\epsfig{file=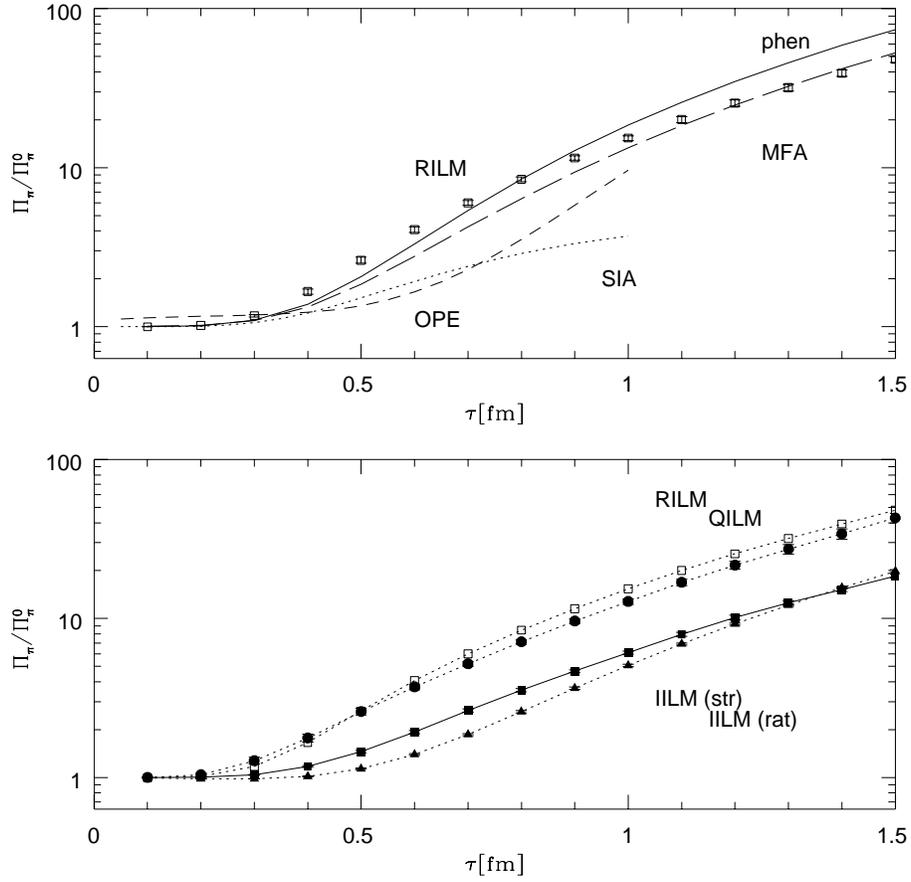, width=5in}
\caption{\label{fig_pi_cor}
Pion correlation function in various approximations and instanton
ensembles. In fig.a) we show the phenomenological expectation
(solid), the OPE (dashed), the single instanton (dash-dotted) and mean 
field approximations (dashed) as well as data in the random instanton 
ensemble. In fig. b) we compare different instanton ensembles,
random (open squares), quenched (circles) and interacting
(streamline: solid squares, ratio ansatz solid triangles).}
\end{figure}

\begin{figure}
\epsfig{file=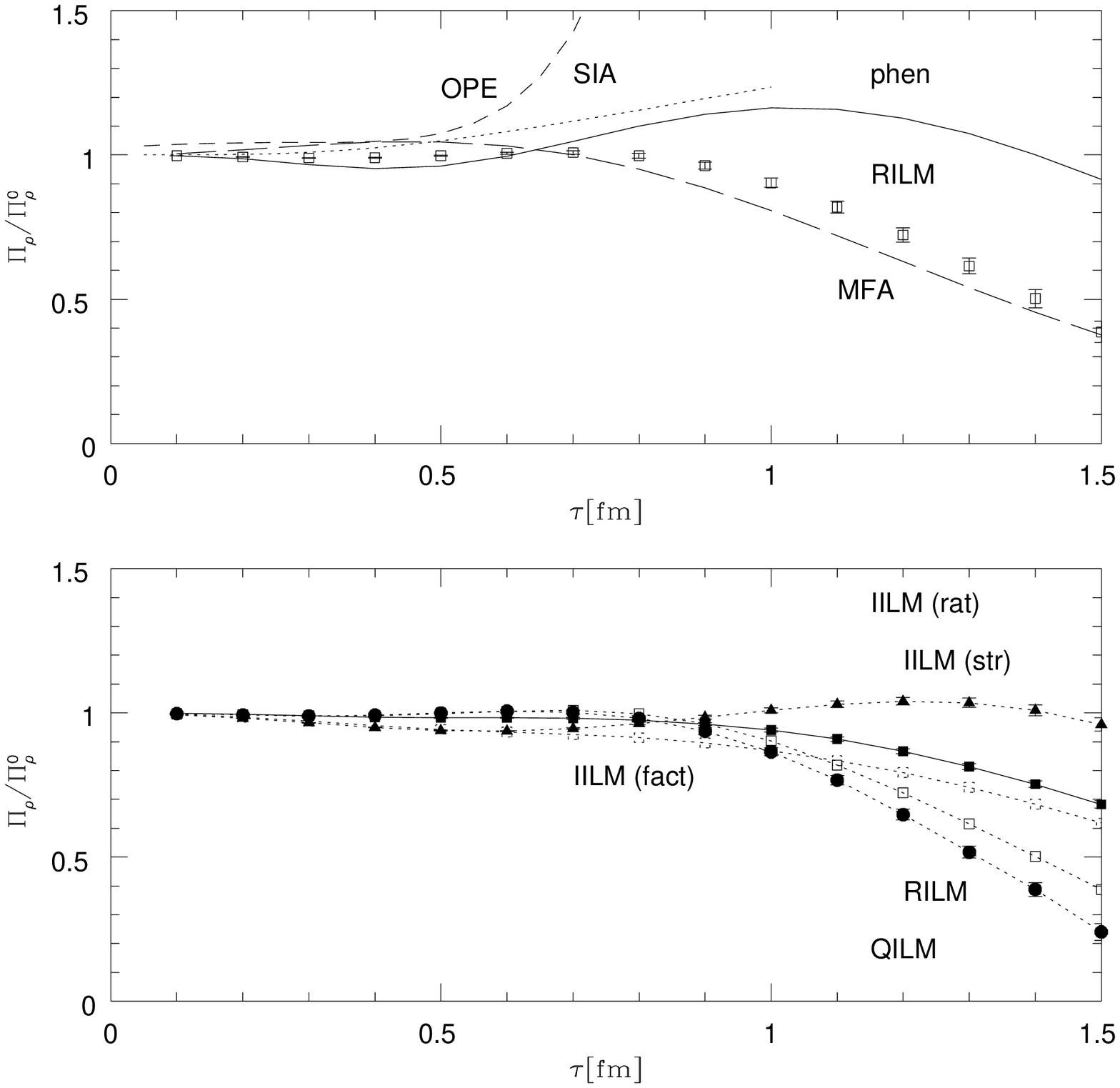, width=5in}
\caption{\label{fig_rho_cor}
Rho meson correlation functions. The various curves and data sets
are labeled as in in Fig.\ref{fig_pi_cor}. The dashed squares show
the non-interacting part of the rho meson correlator in the interacting
ensemble.} 
\end{figure}

In Fig.\ref{fig_pi_cor} and \ref{fig_rho_cor}
  one can see the correlators in $\pi$ and $\rho$ channels.
The functions themselves are strongly falling, and for 
better understanding it is useful to normalize them to ``parton model" 
(or zeroth-order) perturbative diagrams describing propagation of
non-interacting quarks. That is why all figures go to 1 at small x:
it is just 
due to the asymptotic freedom.

  There is no place here to describe all the details.
 Note first
how different the phenomenological lines are: in the pion channel there is strong deviation from 1
already at small distances $x\sim 1/3$  fm or so, while in the rho case
it is close to 1 up to huge $x\sim 1.4$ fm. Why is that?  

Such type of questions were on my mind since the end of 70's, when the
appearance of QCD sum rules  put those under scrutiny. The rho
channels and the like were treated well \cite{SVZ} by including only average fields,
the ``condensates", but for pseudoscalar and scalar channels this approach has
failed completely. It became obvious to me around 1980 that the missing large
effect has to be due to instantons.

The qualitative idea is very simple: 't Hooft interaction has quarks of one chirality
(say left) and anti-quarks only of the other (right) because this is the only
zero modes fermions have in the instanton field. It means large effects in
scalar-pseudoscalar channels (as observed) and no effect in vector-axial
case. Quark mixing pattern (vectors like $\omega$ and $\phi$ are weakly mixed,
while pseudoscalars form combinations  known as $\eta,\eta'$ which are nearly
ideal SU(3)octet and singlet members) supported the same idea. I have calculated
the first-order corrections to the correlators: I got the
 correct signs for pions (attraction) and
the opposite one for $\eta'$ (repulsion). These correlation function, plus
chiral condensates I calculated, has allowed me in 1982 \cite{Shu_82}  
to fix (with some confidence) two main parameters of the instanton
ensemble (\ref{param}).

We will return to their discussion below, and now let us conclude 
with the correlation functions shown in Figs above. The points 
correspond to our first calculation \cite{SV_93}
 using randomly placed
instantons with such parameters \footnote{This simple model is known
  as random instanton liquid model, or RILM.}. In spite of admittedly
quite primitive approach, the points
qualitatively follow these two curves 
(and many others - see the original paper!) well,
reproducing even such details as amazing cancellations of all corrections in the vector channel
mentioned above. So, with only 2 numbers (\ref{param}) and a primitive
model,
 one can  
  calculate the major objects of non-perturbative QCD, 
 the correlation functions, and 
deriving from them  hadronic masses (and much more!) with quite decent
accuracy. We have immediately seen that we are on the right track, and further
work was not disappointing.

There is no place here to discuss its results in details. Let me only mention
that among the $\sim 
40$ correlation functions calculated in the random ensemble, only the $\eta'$ 
(and its U(1) partner, the isovector-scalar) were found to behave wrongly:
 The correlation function decreases very rapidly and becomes 
negative at $x\sim 0.4$ fm. This behavior is incompatible with a normal 
spectral representation. The interaction in the random ensemble was too 
repulsive, and the model ``over-explains" the $U(1)_A$ anomaly. 

In the meantime we \cite{SS_96} have developed the Interacting
Instanton Liquid Model (IILM). Its main element of its partition function
is the fermionic determinant done in zero-mode approximation. It means that
all orders in 't Hooft interaction.
The results obtained in the IILM ensembles 
do not have this problem. Dynamically it is
cased by  correlations between 
instantons and anti-instantons (or the topological charge screening). The single 
instanton contribution is repulsive, but the contribution from pairs is 
attractive. Only if correlations among instantons and 
anti-instantons are sufficiently strong, the correlators are prevented from 
becoming negative.  

Later a very similar situation was found in lattice simulations: the so called ``quenched"
calculations (fermionic determinant ignored) have produce reasonably-looking results for many channels,
but not in the $\eta'$ one. It was realized once again, that
in order to have the QCD vacuum right
one really needs to include the effect of dynamical (and 
rather light) quarks!

 Lattice calculations done later has basically confirmed these
 correlators,
especially in baryonic channels.
Of course lattice people can actually do much more: to hunt for instantons themselves, calculate their density and sizes, correlate them with quark condensate or
hadronic propagators, etc. Amazingly, most the quark condensate states (eigenvalues of the Dirac operator with small eigenvalues) were indeed found to be dominated by
instantons, see \cite{Negele_review}.

\begin{figure}[ht]
\epsfxsize=4.in
\centerline{\epsffile{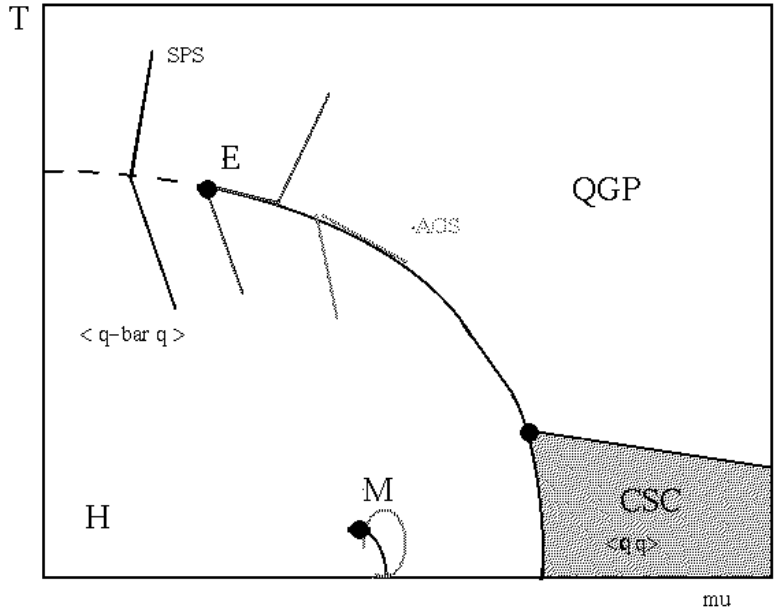}}
\vskip -0.05in
\caption[]{
 \label{fig_phases}
Schematic phase diagram of QCD phases as a function of temperature T
and baryonic chemical potential $\mu$,
as we understand it today. The phases denoted by H,QGP and
CSC
are hadronic, quark-gluon plasma and color superconductor,
respectively.
The dashed line indicate strong cross over, E the endpoint of the 1-st
order transition, M is the endpoint of another 1-st order transition,
between liquid and gas phases of nuclear matter. Few trajectorries
covered in heavy ion experiments are also schematically indicated.
}
\end{figure}

\section{  The phases of QCD}
 The QCD under normal conditions is in chirally asymmetric confining phase
we are so familiar with: but in the so called extreme conditions it turns to
quite different phases. There are at least three different directions 
in which one expect three different phase transitions. (i) At high temperature
$T=T_c\approx 150 MeV$  it undergoes transition to the so called Quark-Gluon 
Plasma (QGP) phase, in which there are no condensates and color interaction is
screened \cite{Shu_78} rather than confined.  (ii) At high density and low T
 it is 
believed to be Color Superconductor (CSC), in which color symmetry is broken
by  diquark condensates induced by instantons\cite{RSSV,ARW}. (iii) At sufficiently large number of flavors $N_f>N_f^c$ it should eventually become chirally symmetric de-confining 
conformal phase.

  The map (for the first two) is shown in Fig.\ref{fig_phases}. We have
  also
shown few schematic trajectories of excited matter, as it expands and
cools
in heavy ion collisions. One may at least see from those that CSC
phase is unfortunately not relevant for them, and so 
we will discuss mostly the high-T direction in this section. Let me only
add few words about the others. 

All transitions are believed to be driven
by instantons. In particularly,
the diquark Cooper pairs are also produced by the same
't Hooft
vertex,  only Fiertz transformed to diquark channels\footnote{The
  perturbative one-gluon exchange also leads to superconductivity, but
  it is significantly weaker in the transition region and can in first
  approximation be ignored.}.
 The most bound diquark
should be the one with  spin and isospin zero (ud), and its condensate is the
largest, reaching the magnitude of about 100 MeV. Other condensates (e.g. us,ds) are smaller, and there are also smaller $\bar q q$ ones, indicated that
at small T and high density the  chiral symmetry is not restored.

The large $N_f$ direction is less studied. At least one reason for that transition is a tendency of instantons and anti-instantons to be bound by the
increasing number of fermion lines connecting them, till finally the ``instanton liquid" is gone and only finite pieces with zero topology (or neutral
``molecules") are left. Calculations in IILM \cite{SS_96} have found  that 
 $N_f^c=5$.

Now we return to the non-zero $T$ case.
Recall that  it is incorporated in quantum field theory
in a very simple way: the Euclidean time $\tau$ is limited by a period $1/T$,
the so called Matsubara time. The instanton solution with periodic boundary 
conditions, called caloron, is well known.
Fermionic  $anti-$periodic zero mode can also be
found. 
\be 
\label{zm_T} 
\psi_i^a &=& \frac{1}{2\sqrt{2}\pi\rho} \sqrt{\Pi(x)}\partial_\mu \left( 
  \frac{\Phi(x)}{\Pi(x)}\right) \left(\frac{1-\gamma_5}{2}\gamma_\mu 
  \right)_{ij} \epsilon_{aj}\, , 
\ee 
where $\Phi(x)=(\Pi(x)-1)\cos(\pi\tau/\beta)/\cosh(\pi r/\beta)$. Note that the
zero-mode wave function  shows exponential decay $\exp(-\pi rT)$ in the 
spatial direction, but oscillates in $\tau$.
So if instantons are like atoms with the quark zero mode as a wave function, 
finite $T$ compresses their special extension and enhances the temporal one. 
(It looks like ``pencil-like" atoms in a very strong magnetic field.) That radically change 
their interactions, which are only strong if instantons are interacting along 
the time direction. In particular, a pair of such type can be formed,
 connected 
to themselves by periodicity.

The main finding in IILM at finite $T$ is that the chiral phase transition is 
actually driven by a {\em rearrangement}\cite{IS_94} of the 
ensemble into a set of instanton anti-instanton ``molecules''\footnote{Note a 
similarity to the Kosterlitz-Thouless transition in the $O(2)$ spin model in 
two dimensions: again one has paired topological objects, vortices in one phase
and random liquid in another. The high and low-temperature phase change places, 
though.}. Recently the details of this mechanism were significantly clarified, 
both by numerical simulations\cite{SS_96} and analytic studies\cite{SV_96}.
   
 At {\em sufficiently high} $T$ new non-perturbative
saddle point appears,  corresponding to a configuration in which the
centers are at the same spatial point, but separated by half a Matsubara box in time $\Delta\tau=1/(2T)$,  the most symmetric orientation of the instanton 
anti-instanton pair on the Matsubara torus. The effect is largest when 
the molecule exactly fits onto the torus, i.e. $4\rho\simeq 1/T$. Using the 
standard size $\rho\simeq 0.33$ fm, one gets $T_c\simeq 150$ MeV,  the 
transition temperature found on the lattice.

In a series of IILM numerical simulations\cite{SS_96} it was found that this 
transition indeed goes as expected, with molecules 
driving the transition. Furthermore, many thermodynamic parameters, the spectra of the
Dirac operator, the evolution of the quark condensate and susceptibilities were
calculated\cite{SS_96,SV_96}, with results surprisingly consistent with 
available lattice data. 
The effect of molecules on the effective interaction between quarks at high 
temperature can be described by the following effective Lagrangian
\be 
\label{lmol} 
 {\cal L}_{\rm mol\,sym}&=& G 
     \left\{ \frac{2}{N_c^2}\left[ 
     (\bar\psi\tau^a\psi)^2-(\bar\psi\tau^a\gamma_5\psi)^2 
      \right]\right. \\ 
     & & - \;\,\frac{1}{2N_c^2}\left. \left[ 
     (\bar\psi\tau^a\gamma_\mu\psi)^2+(\bar\psi\tau^a\gamma_\mu\gamma_5 
     \psi)^2 \right] + \frac{2}{N_c^2} 
%     (\bar\psi\gamma_\mu\gamma_5\psi)^2 \right\} + {\cal L}_8, \nonumber
     (\bar\psi\gamma_\mu\gamma_5\psi)^2 \right\} + \cdots, \nonumber
\ee 
with the coupling constant 
\be 
 G &=& \int d\rho_1 d\rho_2\,\frac{n(\rho_1,\rho_2)}{8T_{\bar I I}^2}(2\pi
\rho_1)^2(2\pi\rho_2)^2\, .
\label{gmol} 
\ee 
Here, $n(\rho_1,\rho_2)$ is the tunneling probability for the $\bar I I$ pair 
and $T_{\bar I I}$ is the corresponding overlap matrix element. $\tau^a$ is a 
four-vector with components $(\vec\tau,1)$.

  There are qualitative things we would like to point out. First,
 some spin-zero 
 states (especially pions and its chiral partner sigma) retain
significant attraction even above $T_c$, and are even likely to
survive the 
phase transition as a bound (but not-Goldstone!) state. 
Second (and maybe relevant for what follows) is that ``molecules" generate
attractive forces in vector channels as well, which were absent below $T_c$.

%%%%%%%%%%%%%% this part to be modified %%%%%%%%%%%%%%%%%%%%%%

\section{  The ``Little Bang'' versus the Big Bang  }
 In general, the field of high energy heavy ion
 collisions
 is now among the most rapidly
developing fields of physics. It is fun to notice its parallels to
cosmology, which goes beyond methodic similarities to amusing
parallels in timing of some recent achievements.

 Appearing at the intersection of high energy
and nuclear physics, these studies are now mostly carried out at CERN SPS
(E=200 GeV*A) and Brookhaven
AGS (E=2-11 GeV*A). Their main goal is production of hot/dense
hadronic matter with the energy density of the order of few $GeV/fm^3$
and study of its properties. 
Especially interesting are $early$ stages
of the collisions, in which
theory predict existence of the QCD phase
transition into a new phase, called the Quark-Gluon Plasma (QGP)
\cite{Shu_78}. The Big Bang of course proceed via T axis in Fig.\ref{fig_phases} since the
baryonic density is tiny. For comparison we have indicated schematic
location and shape of some paths corresponding to current heavy ion 
experiments. 

 The first obvious connection between
the ``Little Bangs" created in these collisions
and cosmological ``Big Bang" is that both
are violent explosions. Expansion of the created
hadronic fireball  approximately follow the Hubble law, although anisotropic
ones. The $final$  velocities of collective motion in both cases have
been
a matter of debates 3-4 years ago, but now are believed to be reasonably
well known. (The main problem here is of course  the reliable
 separation between  
directed collective motion and chaotic thermal one.)
For central heaviest ions the mean transverse velocity reaches about
1/2 c, and so not only longitudinal but also transverse explosion is
relativistic.
 
The second important point 
is that (also in both cases) the underlying history
of matter acceleration, which led to this final velocity,
 remains subject to theoretical speculations.
In ion collisions this was determined by the Equation of State (EoS),
which is believed to be very soft near the QCD transition,
and to find this fact would be very important\footnote{One way to do it,
  is to measure accurately the energy/centrality dependence 
of the collective flow.}.  Experimentally the problem is related with
the fact that observed hadrons
(like microwave  cosmic photons) are seen at 
 their {\it freeze-out stage},
the moment of last interaction. In order to look deeper, one uses rare
particles with smaller cross sections, such as $\Omega^-$ hyperons,
which decouple earlier. Indeed, those show  much smaller
flow. (Cosmologists
solve the problem by looking at very distant Galaxies, 
which also tell you about a
velocity at earlier times.) 

The third type of comparison I would like to make here deals with
the issue of fluctuations.  Very accurate and
difficult measurements of the microwave background anisotropy were made, and
have found (apart of trivial dipole component)
 a trace ($\sim 10^{-5}$) of it originated from plasma oscillations  
at photon freeze-out time. It is seen as some structure with angular
momentum
$l\sim 100$. In heavy ion collisions a similar work is in progress.
The dipole and quadrupole components in azimuthal angle are measured
and are being analyzed: those come from asymuthal asymmetry of initial
conditions, for non-central collisions. We do not yet see reliable
signals of higher harmonics, but I think there is a chance to see
 eventually `` frozen plasma oscillations'' in this case as
 well. After all, we have millions of events, while 
The Big Bang people are restricted to  only one!

 There is no place here to describe heavy ion physics in any details.
 Let me just indicated where we stand now.
 In some respects the heavy ion program was very successful.
Let me  mention one major
issue, which I think by now is pretty much resolved. It is the question
(asked by numerous sceptic s over the years) whether the system produced in
heavy ion collisions is or is not large enough to be treated as a macroscopic
one. The question has been answered positively, as evidences for
  ``flowing"
 locally equilibrated hadronic matter became more and more convincing.
Recent data on event-per-event fluctuations from CERN NA49
\cite{roland} have 
also demonstrated, that in all quantities measured so far the event-by-event
distributions show narrow Gaussian-type histograms, valid for few orders
of magnitude without any visible ``tails". It is very different from
how pp data look like, or from a superposition of NN collisions.
It shows that we are in fact studying here an excited hadronic
system which is very different from that produced in 
pp collisions, and in fact a much simpler one\footnote{I of course mention
it in order to tease high energy physicists, who are proud of studying
simple fundamental topics only, blaming nuclear physicist of usually dealing
with a complicated mess. In this case it is the other way around, due to
significant simplifications arising in the macroscopic limit. } to describe.

  On the other hand, its main goal -- demonstration of the 
QCD phase transition line or presence  of QGP --
is not yet reached. From many observables we see that we definitely
are at the conditions which are the edge or even beyond the
transition. The strong interaction in the
   system, as it expands and cools, erases most of the traces of the
   dense stage.
\section{    Dilepton production in heavy ion collisions:  theoretical considerations   }
   One possible way to study the earlier
 stages  is to look   for ``penetrating probes", production of
$dileptons$ and $photons$. (Another possibility is to look
   for signals which are $accumulated$ during the  evolution:
the well known examples include excessive production of  
strangeness  or charmonium suppression,see lecture by D.Kharzeev).
  Dileptons, unlike secondary hadrons, 
are produced at relatively early stages of the collisions. 
Therefore we expect to see that
hadronic properties 
 are $modified$ in hot/high density hadronic matter, and at high
enough
collision energy we expect to observe the radiation from QGP.

Theoretical calculation of dilepton production is usually made in two steps:
the first is the determination of the {\it production rate} in
equilibrium matter, the second (to which I would not go at all)
is the space-time integration over expansion
of the matter during heavy ion collisions.

The simplest (I would call them 0-th approximation) models for
dilepton production rate are based on well-tested processes:
 (i)  the usual $\pi\pi$ annihilation in hadronic phase at small T , and 
(ii)  it is $\bar q q $ annihilation in QGP for large T. (The second  process is
similar to familiar Drell-Yan process, only in thermal ensemble.)  
those two basic processes can be included in 
 the ``standard rate'' formula:

\begin{eqnarray}
{dR\over d^4q} = {\alpha^2 \over 48\pi^4} F exp{-({q_0\over T})}
\end{eqnarray}

where the rate R is counted per volume per time, q is 4-momentum of
the virtual photon ($q^2=M_{e+e-}=M^2$),
$F$  is 
the usual pion form-factor in the pion gas,
which can be written in standard vector-dominance form. 
In QGP F is a constant, up to small corrections, and we will use this
``partonic'' rate below as our ``standard canle'', the comparison benchmark. 
\begin{eqnarray}
\label{ff}
F = \cases{F_H \buildrel \rm def \over = {m_\rho^4\over[(m_\rho^2 -
    M^2)^2 + m_\rho^2\Gamma_\rho^2]}, &(Hadronic)\cr 
F_Q \buildrel \rm def \over = 12 \sum_q e^2_q\left(1+{2m_q^2\over
  M^2}\right) \left(1-{4m_q^2\over M^2}\right)^{1\over 2}, &(QGP)\cr}
\end{eqnarray}
is a constant in QGP

The ``1-st approximation'' models try to describe what happens in between
these two limits, especially in hadronic matter below but close
to the transition.
It is based on the notion of ``meson modification" in matter.
 We know from standard nuclear physics that the $nucleon$ properties such as
effective mass are modified in nuclear
  matter. There are of course many other examples in condense matter physics,
in which the atomic states are shifted/broadened by the medium. 
So the modification of vector $in-matter$ spectral density 
was usually discussed in terms of shifts of the 
{\it vector meson masses}.
 At low density  one can relate modification of mesons (e.g. of $\rho$)
 to 
the $\pi\rho$ and $N\rho$ forward scattering amplitudes, or
momentum-dependent optical ``potentials'', which predict
relatively modest  shift of $m_\rho$ downward and some broadening.
When matter is no longer dilute and modifications are
no longer small, one should do some re-summations.
As good example of such kind or work let me mention
Wambach-Chanfray-Rapp approach \cite{CRW}, in which very strong
broadening of $\rho$ meson was predicted, based on properties of
$\rho-N$ interaction/resonances. One can also understand this
broadening
as due to mixing between $\rho$ with 
excitations of the lowest baryon resonances,
such as $N^*(1520)N^{-1}$ (resonance plus a  nucleon hole).

  Connecting  hadronic masses to chiral breaking and in particular
to $<\bar q q>$ has lead to the idea that 
$all$ hadronic masses were predicted to vanish  
at $T\rightarrow T_c$. 
 This reasoning has
culminated in the so called
  {\it Brown-Rho scaling} \cite{BR}, according to which
all hadronic dimensional quantities get their scale from $<\bar q q>$ and
 therefore 
\be {m(T) \over m(0)} = { (<\bar q q(T)> \over <\bar q q> })^p \ee
where p is some power (e.g. dimensional 1/3).
Finite T/density QCD sum rules (see e.g.\cite{Hatsuda} and references therein)
also relate hadronic properties to the quark condensate $<\bar q q>$, and
therefore they predict the power p=1.

 One can also find explicit analytic example
of similar behaviour near the phase boundary between chirally
asymmetric and comformal phases of QCD: in this case hadronic scale
driven
by $<\bar q q(T)>$ can be many orders of magnitude smaller than the
basic
``partonic'' scale of the theory $\Lambda$.
On the other hand, at the finite T phase transition
the ``instanton molecules'' mentioned above
may  be important reason for the deviation from such scaling,
 because they generate new interaction in the vector
channel,  unrelated to  $<\bar q q>$.

 Further development along such lines lead to 
what I would call the 2-nd approximation models such as 
 Li-Ko-Brown model\cite{LKB} in which all
vector meson masses are shifted proportional to mesonic density
with Walecka-type mean field. Those models were implemented as
codes, self-consistently describing the evolution of matter and
the mean field. 
%%%%%%%%%%%%%%%%%%%%%%%%%%%%%%%%%%%%%%%%%%%%%%%%%
\begin{figure}[ht]
\begin{center}
\leavevmode
\epsfxsize=5cm
\vskip -0.5in
\includegraphics[width=3.in]{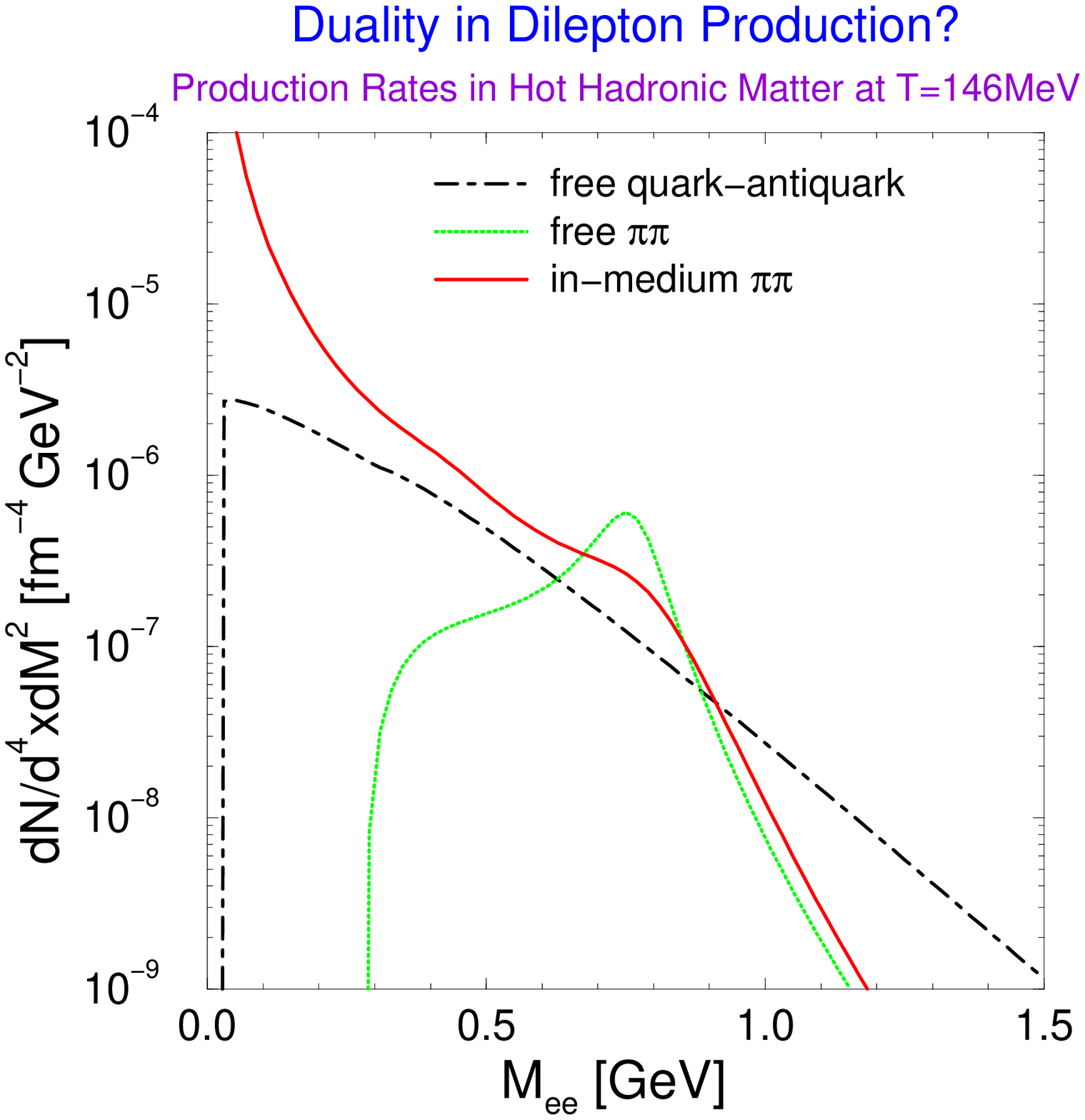}
\vskip -0.5in
\includegraphics[width=3.in]{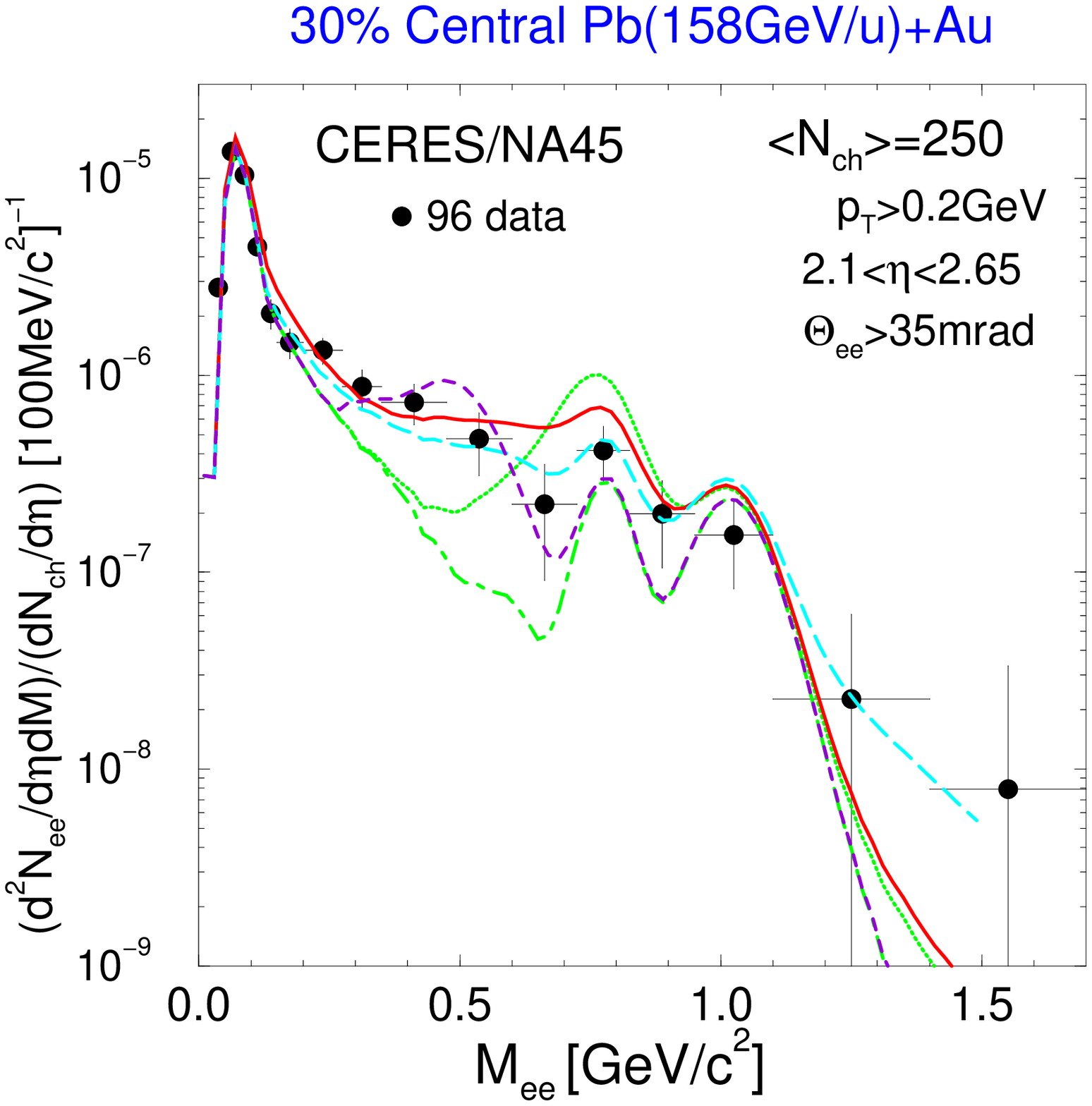}
\vskip -0.5in
\end{center}
\caption{\label{dileptons}
(a) Comparison of dilepton production rates: thermal pion gas,
``partonic''
(dash-dotted) ``realistic'' one (from Rapp et al). 
(b) Comparison of CERES 96 data for mass spectrum of the observed
dileptons with several theoretical calculations:  no
in-matter
production (dash-dotted), no in-matter modification (solid with
$\rho/\omega$ peak), the
Brown-Rho scaling (dashed with a peak at M$\approx$.5 GeV), hadronic
rho widening (solid) and pure ``partonic'' rate (dashed).}
\end{figure}
%%%%%%%%%%%%%%%%%%%%%%%%%%%%%%%%%%%%%%%%%%%%%%%%%%%%%%%%%%%%%%

Let us see how these models compare. In Fig.\ref{dileptons}(a)  one can see  
the rates. The pion annihilation curve show the standard $\rho$  peak,
while for example the ``realistic'' curve including in-matter effects
according to Rapp et al\footnote{The corresponding theory is subject
  to many tests and incorporate a lot of knowledge about cross
  sections and $\rho-N$ resonances, and its details keep changing.
 I am greatful to Ralf Rapp who
  provided this updated figures prior to publication.}. 
The widening of $\rho$ plus Boltzmann factor which emphasizes small
masses have changed this curve significantly.
Note also one
striking fact: for the most important
masses $M=.3-.6 GeV$ the ``realistic'' curve obtained in a complicated
calculation  
is  not so far from the ``partonic'' one, which would correspond to
just ideal gas of quarks and anti-quarks.

  Another way to demonstrate to the students here that
 the matter is by no means settled, even qualitatively, is to mention
 quite opposite suggestions about hadronic properties close to (or
 even above) $T_c$ which can be
found in literature. For example, effective Lagrangians lead
to a prediction of $rising$ $m_\rho(T)$ \cite{Pisarski_95}, moving by
$T_c$ 
about half way toward the mass of its chiral partner $a_1$.
Do we have any firm theoretical benchmark here?

 One definite thing is  that chiral restoration demands that
vector spectral density should become identical to that of the $axial$
current. Indeed,
the only difference between them is related with chirality-flipping terms.
More specific form of this statement can be written as a set of Weinberg-type
sum rules \cite{KS},
 related the certain moments of the difference with a particular
chirally-odd quantities like $<\bar q q>,f_\pi$.

However,  actually we expect more from chiral restoration than just shifting
$\rho$ and $a_1$ to the same point, and thus eliminating the difference.
The high-T phase is expected to be the QGP, with only perturbatively 
interacting
quarks and gluons. If so, one may expect that delta-functions (resonances) would be effectively gone,
and the threshold energy $E_0$ move down, somewhere to twice perturbative quark
mass in QGP.
 This is basically what CERN dilepton experiments such as CERES have indeed
indicated (And that is why I have found this subject to be worth presented
at this prestigious school!) Very roughly speaking, the data are consistent
with nearly ``partonic" production rate, and not only at very high
density/temperatures in QGP, but already in hadronic phase
close to $T_c$.

What is still missing in theory, in my opinion 
(the ``3-ed generation models") is
models based on more fundamental level, 
capable of explanation of the modified basic interactions between quarks.
Those should be able not only predict changes of masses, but also
of condensate themselves and other parameters of the spectral density.
I think the most important issue is actually the modification
of the  ``duality scale'', the
threshold" $E_0$, above which parton results are dual to hadronic
calculations.
In vacuum this parameter is around the mass of the second resonance,
$E_0=1.4-1.6$ GeV, related to
``gluon condensate'' by the QCD sum rules.
What is puzzling here, is that
  at $T\approx T_c$ we have in these collisions this parameter $E_0$
 seem to become
very small, not larger than 
about .3 GeV or so, while the gluon condensate cannot
dramatically change from T=0.

%%%%%%%%%%%%%%%% 

\section{    Dilepton production in heavy ion collisions:   experiments}
 There are three dilepton
 experiments  at CERN SPS:
(i) CERES (NA45), which study the low mass  (M=0-1. GeV) $e^+e^-$ pairs,
(ii) HELIOS-3, which study medium mass $\mu^+\mu^-$''  M=1-2 GeV,  
and (iii) NA38/50 concentrated on high mass $\mu^+\mu^-$. All three 
see quite significant enhancement over ``standard sources'', ranging
from factor 5 at CERES (in some kinematic region) to about 3 at
NA38/50\footnote{In the ``very high" mass region, M>3 GeV,
dilepton production is well described by simple partonic Drell-Yan process.} at M=2-3 GeV. These numbers are maximal,
corresponding to the most central heavy ion collisions, like Pb Au.
 Let me also mention that old Bevalac dilepton energy experiment, at E$\sim$ 1
Gev*A also found strong enhancement: these results would be soon tested
at SIS in Darmstadt. 

I have no place to explain that experimentalists have done their homework,
they have measured the dilepton production in pp (or p-Be) and found the
results completely consistent with a ``cocktail'' of known effect, such
as $\pi,\eta$ and
resonance decays. 
 CERES data for heavy (PbAu) ion collisions are shown in Fig.\ref{dileptons}(b): one can clearly 
see that relative to free pion gas annihilation (the no-in-matter-modification
curve) 
there is a
 low-mass $enhacement$, combined with the $deficite$ in $\rho,\omega$
mass region. The curves related with ``shifted'' $\rho$  mass a la
Brown-Rho
or ``realistic'' widened $\rho$ are consistent with the data.
More detailed data (not shown) indicate that these effects 
 exist  only for low $p_t$ dileptons, indicated
that it is indeed an in-matter effect. It  excludes some models
(like that based on P-wave $\rho-N$ resonances.
CERES has been upgraded and the 98 run should provide much better
resolution and signal/background ratio, although comparable statistics.
  Helios-3 data (not shown) are also provide evidences for in-matter 
production, which is consistent with these two theories, or our
benchmark, the
``partonic'' rate.  

At the present level of accuracy, it is only possible to conclude that
the spectral density of in-matter excitations is indeed qualitatively
 different from
in-vacuum one. The $\rho$ peak seem to be  gone, and the spectral
density looks  close to ``partonic'' quark continuum we expect to
see in QGP. It is actually quite puzzling fact,
 since only small fraction of the
 space-time volume contributing to dilepton production is expected to
be QGP above the transition, while most of it should still 
be in hadronic phase. 

Let me at the end mention few  unresolved issues.
One is what happens with $\omega$ peak: it will be answered by CERES
soon. The other is
 whether in-matter modification is the mesonic  or baryonic effect. It could
be tested by going to RHIC (very small baryon/meson ratio) and SIS
(large baryon/meson ratio). Finally,
 the NA38/50 enhancement is not yet analyzed:
it may be due  either to true QGP dileptons or
the enhanced charm production. Both are very exciting possibilities,
maybe a decisive clue to the QGP.

   At the end of this section, let me make some  remark about
status/history of the dilepton experiments in general.
    These kind of experiments are generally much more 
difficult  compared to  measurements of hadronic observables.
In addition to large background which need to be rejected, they
are related with smaller cross sections and are significantly limited
by its statstics/number of runs.
Historically 
 dilepton experiments have a tendency to came too late.  At Berkeley
BEVALAC the DLS spectrometer had so limited
 number of runs  that its results 
will probably be never  understood. The
Brookhaven AGS program had no dilepton experiments
at all. The CERN SPS program has 3 experiments mentioned above, and those 
have produce exciting
data. However, they all have one run a year, for few weeks. The data
are  still statistics limited and  had so few runs that
one cannot trace dependence on even such major parameters as collision
energy or atomic number. And the end of SPS heavy ion program may be already
at sight. 

 Next year a dedicated relativistic heavy ion collider (RHIC)
 would start its operation
at Brookhaven. It would collide 100 GeV per nucleon beams of Au nuclei with
each other, and among its detectors one (PHENIX) is mostly devoted to
electron, muon and photon physics. 
Heavy ion program is also a part of LHC project, with one
common European heavy-ion-oriented detector ALICE. With those facilities, we
 should be able
to penetrate deeply into domain of the QGP phase.

\section{        Conclusions and discussion}

  I have argued that $instantons$ 
dominate nearly all aspects of light quark physics in QCD.
The instanton  ensemble is dense enough to be in a disordered
phase, so that their zero
   modes are collectivized into a ``zero mode zone"
with small Dirac eigenvalues. These states form
  the non-zero quark condensate.
As shown both on the lattice and in the instanton models, these states
dominate also light quark propagators and hadronic
correlation functions at large distances.

 At    temperature $T>T_c\sim 160 MeV$ the chiral symmetry  
gets restored because random   instanton
ensemble breaks  into finite clusters.
The instanton simulations show that these clusters are
particularly structured instanton- anti-instanton pairs or ``molecules".
Lattice simulations had only partially supported this scenario:
larger-volume and smaller-mass simulations are needed to 
clarify this issue.

At low T and high density there is another QCD phase, known as
color superconductor. In many respects it is closer to electroweak
theory, because the colored diquarks condense and break color symmetry,
like Higgs particles. It is interesting, that with the inclusion of
strange quarks one finds a very particular hierarchy of the condensates, 
including  chirally 
asymmetric $<\bar q q >$.

  In our discussion of the QCD vacuum/hadronic structure we have
pointed out a striking difference between the spin zero channels
(such as $\pi,\sigma,\eta'$) and vector ($\rho,\omega,\phi,a1$)
ones. The former show strong deviations from parton behavior at small
distances, the latter do not. 
  Now one can ask what happens with inter-quark interaction as
  $T$ is approaching  $T_c$ from below. Theoretically one finds significant
changes in masses of all scalars, but those is difficult
to observe. However vector spectral density is directly measurable 
in dilepton experiments.

 All three dilepton experiments at CERN: CERES, 
 HELIOS-3 and  NA38/50  see strong enhancements relative to ``trivial
 sources'', especially at small  $p_t$. It is clear from the data that
 in-matter production of dileptons is significant. Furthermore, a
modification (or maybe even complete melting) of $\rho$ resonance
 seem to  take place.
  Some success in phenomenological treatment with shifted
   masses and modified widths was achieved at hadronic level. 
However
at more fundamental quark level we still do not quite understand what
is the effective interaction between quarks in the vector  channel
at $T\approx T_c$. It looks like
 ``partonic" rate approximately describes the production at $T\approx T_c$
in the whole of  domain of hot/dense matter created at SPS, up to very
small masses of the order of .3 GeV. This is rather surprising.
If so, it which would imply that we see some ``premature" chiral symmetry
restoration, even before the QGP phase has really become dominant.
Much more work, both experimental and theoretical, is needed to
clarify this outstanding claim.

\section{Acknowledgments}
I am greatful to organizers,
Profs.G't Hooft, G.Veneziano and A.Zichichi, for invitation to
this school.  Over  30 years I had several invitations to
Erice schools, but sadly  it happened so that this 
is the first I  actually used. I am very impressed by its scientific
level,  while the  level of hospitality set a record that can
hardly  be beaten. Let me also mention that
my lecture is based on works done with  collaborators, especially with 
J.Verbaarschot and T. Schaefer, who contributed immensely to progress reported 
in them. My research is partially supported by US DOE.

\end{document}